\documentclass[referee]{mn2e}
\usepackage{graphics}
\title[The 11.3-$\mu$m SiC feature]{The screening of the 11.3-$\mu$m SiC feature by carbonaceous mantles in circumstellar shells}

\author[R. Papoular]{R. Papoular$^{1}$\thanks{E-mail:
papoular@wanadoo.fr}\\
$^{1}$Service d'Astrophysique and Service de Chimie Moleculaire,\\
CEA Saclay, 91191 Gif-s-Yvette, France}
\begin{document}

\date{Accepted . Received ; in original form }

\pagerange{\pageref{firstpage}--\pageref{lastpage}} \pubyear{2002}
   \maketitle
\label{firstpage}
\begin{abstract}
Silicon carbide, a refractory material, condenses near the photospheres of C-rich AGB stars, giving rise to a conspicuous emission feature at 11.3 $\mu$m. In the presence of a stellar wind, the SiC grains are carried outwards to colder regions, where less refractory carbonaceous material can condense, either by itself or in mantles upon SiC grains. Enough carbon can condense on the latter that their specific feature is completely veiled. Thus may be explained a) the coexistence of the SiC feature protruding above a carbonaceous continuum, with a range of contrasts, corresponding to various volume ratios of mantle to core; b) the ultimate disappearance of the 11.3-$\mu$m feature from the interstellar medium, where the mantle has become completely opaque due to the much higher cosmic abundance of carbon.
  \end{abstract}

\begin{keywords}
Core-mantle grain condensation; expanding stellar atmospheres; 11.3-$\mu$m SiC feature
\end{keywords}
%
\section{Introduction}

The 11.3-$\mu$ feature discovered by Hackwell \cite{hac} is observed in absorption towards many C-rich stars and is attributed to small SiC grains condensed from their expanding envelopes. The shape of the feature was later extensively documented by IRAS (InfraRed Astronomical Satellite \cite{iras}, which detected about 538 such features, gathered by the IRAS Team in their Class 4n. Little-Marinin \cite{lit}, Baron et al. \cite{bar}, Papoular \cite{pap88}, Borghesi et al. \cite{bor}, among others, analyzed and classified the spectra in this data bank. This generated an abundant literature, mostly focused on explaining the various shapes and widths of the feature (e.g. Treffers and Cohen \cite{tre}, Goebel et al. \cite{goe}, Dorschner et al \cite{dor}, Kozasa et al. \cite{koz96}) by comparison with laboratory measurements on SiC powders and crystals (e.g. Borghesi et al. \cite{bor}, Papoular et al. \cite{pap98}.

However, not much effort went into understanding the large span of observed feature intensities; its strength relative to the underlying continuum is mostly found to range from $\sim$0.1 to $\sim$3 (Baron et al. \cite{bar}); nor is it understood why this same, conspicuous, feature is never observed in the ISM (interstellar medium), although SiC grains are found and studied in meteorites.

This is precisely the subject of the present work. Its starting point lies in the calculations performed by Kozasa et al. \cite{koz96} with the purpose of showing the similarity of the 11.3-$\mu$m feature profile with the results of Mie computations of spherical grains consisting of a SiC core coated with a carbon mantle, for a range of grain radius and volume fraction of carbon. Also, in an earlier paper, Frenklach et al. \cite{fre} showed experimentally that SiC nucleates at high temperatures and could provide a surface for subsequent carbon condensation.

The present effort was prompted by the observation, in their Fig. 3, that the SiC bump in the spectrum of the composite grain is red-shifted and weakened as the C-fraction increases. It is barely distinguishable when this fraction exceeds 0.9.

Here, we intend to explore the conditions under which such composite grains can grow. It is found that mantles can generally form and become thick enough to completely screen the 11.3-$\mu$m SiC feature. The core-mantle configuration could thus explain the large range of feature intensities and its ultimate disappearance from the ISM.

The usual procedure in the field of grain formation is based on the classical thory of homogeneous condensation. For the expanding circumstellar shells of AGB stars or novae, this must be coupled with mechanical equations describing the action of the forces that drive the stellar atmospheres. It is generally assumed that the radiation pressure exerted by the stellar photons upon the grains accelerates the latter through the gaseous atmosphere. Collisions of the grains with the dominant hydrogen atoms or molecules produce a dragging force which lifts the whole atmosphere upwards, giving rise to the stellar, ``grain-driven", wind.

Among the many authors who developed this model, we relied mainly on Salpeter \cite{sal}, Draine and Salpeter \cite{dra}, Kwok \cite{kwo}, Lefevre \cite{lef}, Deguchi \cite{deg}, Kozasa et al. \cite{koz83} and Gail et al. \cite {gai}, especially for the dynamic part. However, an enormous uncertainty plagues the nucleation rate in the various treatments of homogeneous condensation, namely, several orders of magnitude on the pre-exponential coefficient (see discussion in Deguchi \cite{deg}). Instead, the treatment adopted here considers the number density of grain seeds as a parameter to be specified tentatively at the point of condensation. In effect, its impact on the conclusions is found to be only marginal.

In the following three sections, for the sake of clarity, we treat separately and successively the 
condensation of a single species in a static atmosphere, then its condensation in a grain-driven wind, and, finally, the formation and growth of composite grains in an expanding atmosphere. The last section discusses the IRAS observations of C-rich stars in this context.

\section{The static atmosphere}

Consider a star of mass $M_{*}$, radius $R_{*}$, luminosity $L_{*}$ and surface temperature $T_{*}$. In the absence of wind, its atmosphere is in static thermodynamic equilibrium. Radiative or thermal energy exchanges being assumed negligible, the equilibrium is adiabatic and characterized by the ratio of the specific heats, $\gamma$, in 

$PV^{\gamma}=constant;\,\,\,   TV^{\gamma-1}=constant.$

In this case, the number density of gaseous hydrogen, which is inversely proportional to the specific volume, decreases with the altitude, $h=R-R_{*}$, according to Kennard \cite{ken}

\begin{equation}
n_{H}(r)=n_{H}(1-\frac{\gamma-1}{\gamma}\frac{gh}{RT_{*}})^{\frac{1}{\gamma-1}},
\end{equation}

while

\begin{equation}
T(r)=T_{*}(1-\frac{\gamma-1}{\gamma}\frac{gh}{RT_{*}}),
\end{equation}

where $g=GM_{*}/R_{*}^{2}$ is the local gravity and $R$, the universal gas constant, $8.31\,\, 10^{7}$ erg.deg$^{-1}$.mole$^{-1}$. For a monoatomic gas, $\gamma=5/3$, and it increases with molecular size ( degrees of freedom, in fact) towards 4/3, through 7/5 for diatomics. We shall adopt 3/2 to allow for molecular hydrogen.

Near the star, $n_{H}$ decreases slowly, then very quickly near the ``edge"

\begin{equation}
R_{e}=R_{*}(1+0.43\,\frac{(R_{*}/2\,\,10^{13})(T_{*}/2500)}{M_{*}/2\,\, 10^{33}}),
\end{equation}

in degrees Kelvin, cm, grams.

Equation (2) can then be rewritten

\begin{equation}
T(r)=T_{*}(1-\frac{R-R_{*}}{R_{e}-R_{*}}),
\end{equation}

which, by inversion, gives the condensation radius for a given grain condensation temperature, $T_{c}$:

\begin{equation}
R_{c}=R_{*}(1+0.43\,\frac{(R_{*}/2\,\,10^{13})(T_{*}/2500)}{M_{*}/2\,\, 10^{33}})(1-\frac{T_{c}}{T_{*}}).
\end{equation}

It is instructive to study the growth of grains beyond the point of first condensation, in a static atmosphere, as it displays the essentials of the growth process without the complications occasioned by a wind. Let us assume, for simplicity, that the densities and temperatures are uniform. Let $n_{H}, n_{1}$ and $n_{g}$ be the number densities of hydrogen, of monomers of the condensible material and of the condensed grains, respectively, at a given time. Assume the grain population to be homogeneous, with a common number, $\nu$, of monomers condensed in each grain, corresponding to a common grain radius, $a$. Let $u_{1}$ be the average velocity of thermal agitation of the monomers, and $\sigma_{1}$ their cross-section for sticking together. Since $a \propto \nu^{1/3}$, assume that the cross-section for sticking of a monomer to a grain is $\sigma=\sigma_{1}\nu^{2/3}$, proportional to the grain geometric cross-section. Then, $n_{1}$ decreases in time according to  

\begin{equation}
\dot n_{1}(t)=-c_{1}n_{1}n_{g}\nu^{2/3},
\end{equation}
where the dot designates the first derivative and $c_{1}=\sigma_{1}u_{1}$.

The grain size increases by accretion of monomers but may also do so by coalescence of grains. Since the grain agitation velocity is inversely proportional to the square root of its mass (or volume), we write it as $u_{1}/\nu^{1/2}$. Now, for each collision between two grains, the gain in grain size is $\nu$, which must be shared by $n_{g}$ grains, if homogeneity is to be maintained. Then, 

\begin{equation}
\dot\nu(t)=c_{1}\nu^{2/3}n_{1}+c_{1}\nu^{7/6}n_{g}.
\end{equation}

When accretion has nearly depleted the monomer reservoir, coalescence takes over as $n_{g}\nu$ tends to $n_{0}$, the initial number density of monomers, and expression (7) tends to $c_{1}n_{0}\nu^{1/6}$.

Finally, conservation of matter is ensured by 
\begin{equation}
n_{g}(t)=\frac{n_{0}-n_{1}}{\nu}.
\end{equation}

If the coalescence term in (7) is negligible, $n_{g}$ remains constant because $\dot n_{1}=-n_{g}\dot\nu$ (eq. 6 and 7).

As an example, take SiC as the condensible material. It is known as a highly refractory material, with a measured melting temperature of 3100 K (Touloukian \cite{tou}). Although this does not necessarily mean that it also condenses in circumstellar shells at this temperature, we shall consider that,  for $T_{*}=2500\,$ K, it will condense near the photosphere. The cosmic abundance of Si relative to hydrogen is $\sim10^{-5}$, much less than that of condensible C. We therefore take the former as the abundance of SiC monomers. The monomer has an atomic mass of 40 and the specific gravity of crystalline SiC is 3.2, so $a=1.7\,\,10^{-8}\,\nu^{1/3}$.

The solution of the above system of equations, two of which are 1st-order differential, requires the knowledge of initial values of the corresponding variables. The range of realistic values of $n_{H}$ is determined by observed values of mass loss, $\dot M$ and wind velocity, $V$. As to the initial value of $\nu$, it is obviously 1. The problem lies in the choice of the relative number density of grains, $n_{g}/n_{H}$. This indeterminacy parallels the usual one in classical homogeneous nucleation theory, referred to in the Introduction. Here, we limit ourselves to exploring the impact of this parameter on the final grain size and number density.

Assume, for instance, that the initial number densities of hydrogen and SiC grains are respectively $10^{11}$ and $10$. The solution of the above set of equations then delivers the curves of Fig. 1. Grain growth by accretion is seen to quickly deplete the reservoir of monomers: the fraction of the latter that are captured by grains reaches 1 in less than a year. Grain growth by coalescence becomes apparent afterwards and is accompanied, as expected, by a reduction in grain number density. However, this process is much slower, and does not contribute significantly even after 100 years. This remains true for all situations considered here.

\begin{figure}
\resizebox{\hsize}{!}{\includegraphics{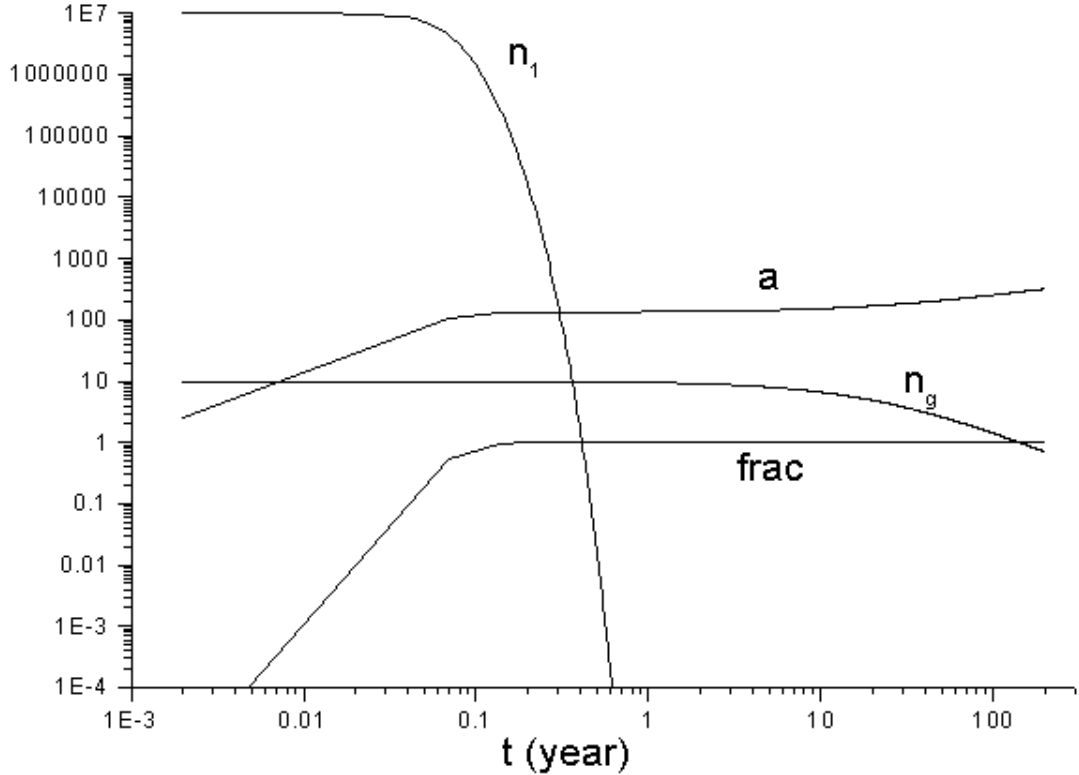}}
\caption[]{The time evolution of monomer and grain local number densities ($n_{1}$ and $n_{g}$ in cm$^{-3}$), grain radius in Angstrom, $a$, and fraction of the monomer reservoir trapped in grains, $frac$, in a static atmosphere, at a point in space where the condensible material is SiC and, initially, $n_{H}=10^{11}$ cm$^{-3}$ and  $n_{g}=10$ cm$^{-3}$. Note grain size increase and grain number density decrease, after 1 y, due to coalescence. Nearly all the condensible material is condensed after about 0.1 y.} 
\end{figure}

If, now, the initial density of grains is reduced to $10^{-4}$, for the same $n_{H}$, the contribution of coalescence is invisible but the final grain radius increases to 0.35 $\mu$ (Fig. 2) from 0.017 in the previous case. 

\begin{figure}
\resizebox{\hsize}{!}{\includegraphics{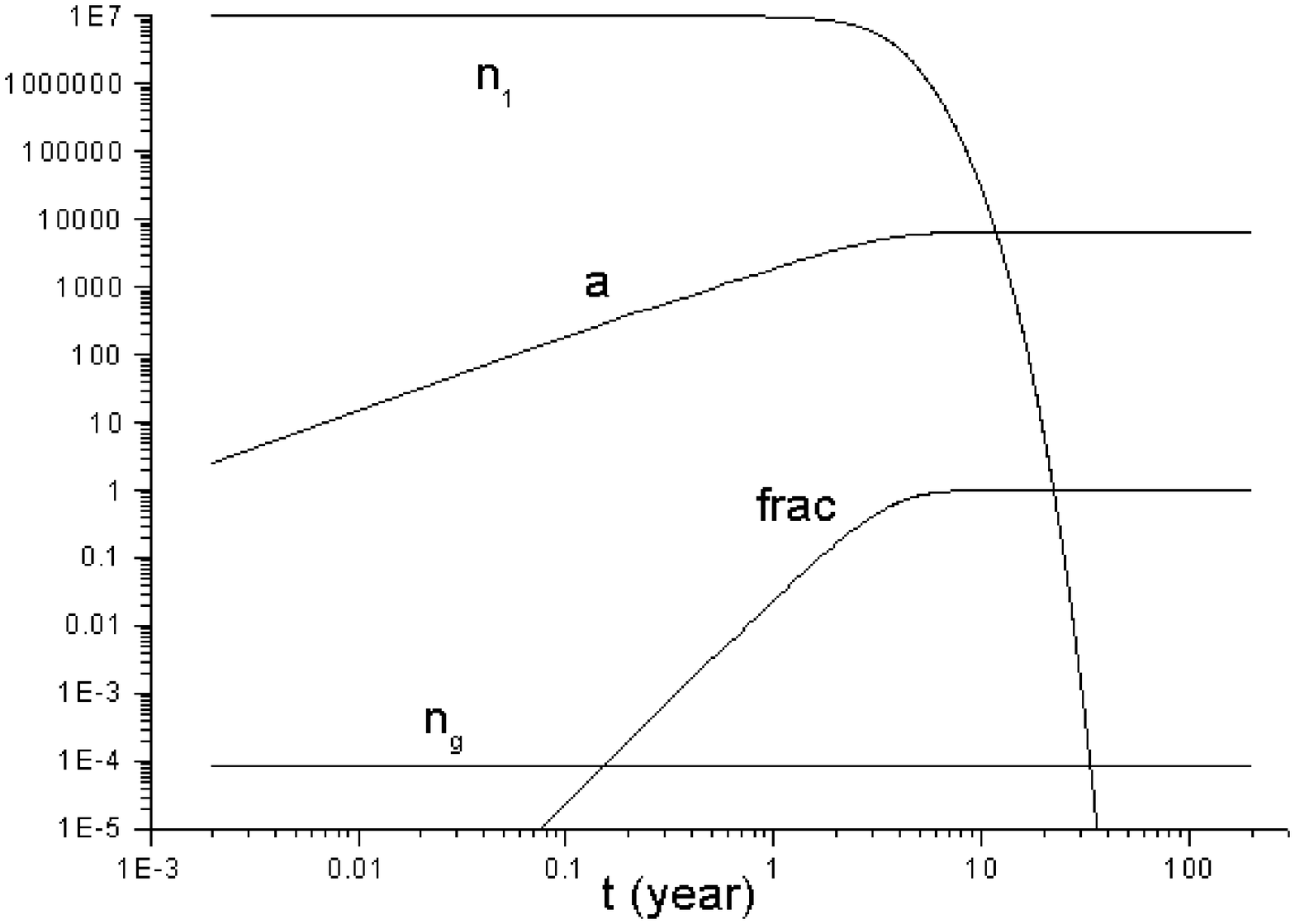}}
\caption[]{Same as Fig. 1, except that, here, the initial nuclei density, $n_{g}$ is 10$^{-4}\,$ cm$^{-3}$. No sign of coalescence observed after 100 y, but the ultimate grain size is much larger; however, this requires a much longer time.} 
\end{figure}

Grain growth and monomer fraction in grains are drawn for intermediate cases in Fig. 3. Note that a factor $10^{7}$ decrease in $n_{g}$ entails only a factor 100 increase in final grain size. This only confirms Deguchi's conclusions \cite{deg}. It is also consistent with the range of the grain sizes deduced from astronomical observations.

\begin{figure}
\resizebox{\hsize}{!}{\includegraphics{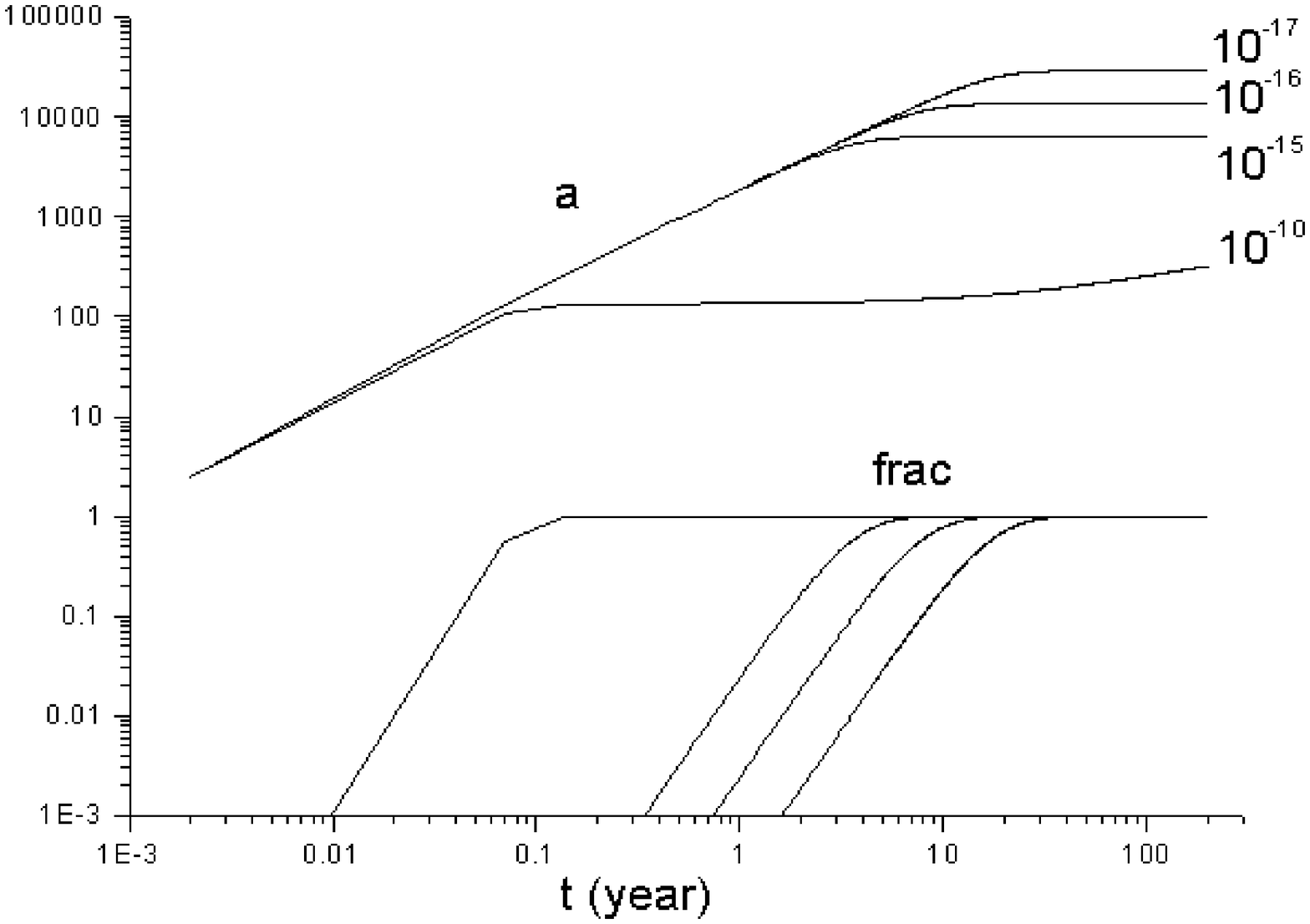}}
\caption[]{Grain size $a$ and condensed fraction $frac$, for a range of initial relative nuclei densities ($n_{g}/n_{H}$).} 
\end{figure}
Since the monomers are quickly depleted, $n_{g}\nu \sim n_{0}$ and a simple relation holds between grain size and initial relative concentration of nuclei:

$a=1.7 \,\,10^{-8}\,\nu^{1/3}=1.7\,\,10^{-8}\, (\frac{n_{0}}{n_{g}})^{1/3}$.

Also, the time required to reach the ultimate grain size increases nearly linearly with the latter.
One may muse that, independent of any theoretical indeterminacy, the inherent turbulence of stellar outflows, as well as the time factor, may cause by themselves the observed grain size dispersion.

Although the atmosphere considered here is static as a whole, condensed grains are accelerated  under radiative pressure from the star. Kwok showed that this dragging force drives the grains forward relative to the gas, with a velocity
\begin{equation}
V_{d}=[\frac{1}{2}[(4K^{2}+u^{4})^{1/2}-u^{2}]]^{1/2},
\end{equation}
where $K=\bar Q_{pr}L_{*}/4\pi c\alpha\rho_{H}r^{2}$, and $\alpha=3/4$. This drift velocity was computed and found to be less than 0.1 cm$^{-1}$, in the present case. This is very slow and may exclude the expulsion of grains in the interstellar medium. However,the pure SiC particles formed in the inner shell will contribute a strong 11.3-$\mu$m feature against the background of the hot photospheric continuum (cf. Baron et. al. \cite{bar}). Density integration along the star direction shows that, for all reasonable photospheric hydrogen densities (less than $10^{12}$ cm$^{-3}$), the optical thickness of condensed SiC remains smaller than 1, justifying the fact that the feature is never observed in absorption.
In the static atmosphere considered in this Section, carbon can also condense, but farther out than SiC, according to eq. (4), because its condensation temperature is distinctly lower. However, the equations above indicate that, with increasing altitude, the gas density decreases much faster than temperature which may hinder any substantial carbonaceous condensation, so no screening mantle is formed on the SiC grains, in general, in static atmospheres.
 
\section{The expanding atmosphere; single grain species}

Consider now the case where the star luminosity and condensible species concentration are high enough for the grains to accelerate the atmosphere upwards. Following Kozasa et al. \cite{koz83}, we write the equation of motion

\begin{equation}
(\frac{V}{u^2}-\frac{1}{V})\frac{dV}{dr}=\frac{2.5}{r}-\frac{GM_{*}}{r^{2}u^{2}}(1-F) 
\end{equation}

where $u$ stands for the velocity of sound in the atmosphere, and the radiation pressure to gravity ratio is

\begin{equation}
F(t)=\frac{n_{g}a^{2}\bar Q_{pr}L_{*}}{4cGM_{*}\rho_{H}}.
\end{equation}

Here, $L_{*}$ is the star luminosity, $c$ is the velocity of light and $\rho_{H}$ the specific density of the hydrogen gas ($1.5*1.7*10^{-24}n_{H}$), in units of cm, g, erg, sec. The dimensionless factor $\bar Q_{pr}$, which characterizes the efficiency of the stellar photons for pushing the grains outwards, depends on the stellar temperature and the properties of the condensible material. Gilman \cite{gil} computed its value, on the basis of Mie's theory, for SiC and graphite, in particular. Based on his work, we take here the ratio $\bar Q_{pr}/a$ to be 1000 cm$^{-1}$ both for SiC and carbon grains.

In eq. 10, $F$ starts from 0 at the condensation point, $T=T_{c}$, and increases as the grains grow along the wind, to reach $F_{max}$ if and when the available monomers can all condense, i.e. when $n{g}\nu=n_{0}$; then

\begin{equation}
F_{max}=3.4\,\mu^{3}f_{at}Q\frac{(L_*/10^{4}L_{\odot})}{M_{*}/M_{\odot}}, 
\end{equation}

where $f_{at}$ is the cosmic abundance of the considered species relative to hydrogen, $\mu=a/\nu^{1/3}$=1.7 for SiC and 1.4 for C, and $Q=\bar Q_{pr}/a$ (cm$^{-1}$). Expansion can only start if and when, along the way, $F$ becomes slightly larger than 1 so the r.h.s. becomes positive. As the term $2.5/r$ is generally negligible in eq. 10, a necessary condition for the wind is $F_{max}>1$. $F_{max}$ is related to the critical stellar luminosity, $L_{cr}$, for a star to drive a wind, as previously defined by Salpeter \cite{sal}, by $F_{max}=L_{*}/L_{cr}$. As $Q$ hardly ever exceeds $10^{5}$ cm$^{-1}$, species less abundant than $10^{-4}$ require quite high stellar luminosities to lift a wind; hence their relative rarity.

Equation (10) implies that the sign reversal occurs at the sonic point, where $V=u$ and both sides are null, so $dV/dr$ is indeterminate. All previous authors signaled the difficulty of finding a stable solution that discribes the flow through this critical point. Here, we skirt round the problem by using the following $\emph{ansatz}$. First, assume the sonic point to coincide with the condensation point. This is justified by the shortness of the time required for condensation, as evidenced by Fig. 1 to 3. This assumption is also borne out by the calculations of Deguchi \cite{deg} and Kozasa et al. \cite{koz83}, at least over a large range of luminosities and mass losses. At any rate, it does not seem to affect notably the subsequent evolution of velocity and grain size; that is enough for present purposes. Since our calculation starts at the critical point, we also impose $F$=1 at that point, which implies that, although the initial grain number density may be chosen arbitrarily, the grain size must be determined correspondingly so as to satisfy that condition on $F$.

Returning to eq.(10), and remembering that $dr=Vdt$, where t is the time, we rewrite it as

\begin{equation}
\dot V=\frac{\frac{GM_{*}}{r^{2}}(F-1)+\frac{2.5u^{2}}{r}}{\frac{1}{u^{2}}\frac{1}{V^{2}}}.
\end{equation}

We also assume that the condensation temperature is independent of mass loss and take the sound velocity to be the local thermal velocity. Then, applying l'Hospital's Rule, we differentiate numerator and denominator at the critical point. Taking into account the swiftness of condensation, the variation of $r$ can be ignored. Some algebra delivers the wind acceleration from

\begin{equation}
(\dot V_{c}/u)^{2}=F_{max}n_{gc}\frac{c_{1}GM_{*}}{2r_{c}^{2}u}.
\end{equation}
 
 On the r.h.s. of eq.(14), $F_{max}n_{gc}c_{1}$ is recognized as the ratio of the maximum radiative force on the grains to the mean free path of monomers against sticking to grains. The physical meaning of this becomes obvious when it is remembered that the dragging force exerted by the photons on the gas via the grains increases as the mean free path decreases (see Kwok \cite{kwo}).

As for the static atmosphere (Sec. 2) we computed the drift velocity of the grains relative to the gas, and found that, in all considered cases, it was negligible compared to V, as shown below. Its neglect in our equations of motion is therefore justified $\emph{a posteriori}$; grains and gas can be considered as locked together in the flow.

With the initial acceleration given by eq.(14) at the starting point $r_{c}$, eq.(10) can be solved to give the growth of grains in time and space. Because of the flow and consequent expansion, all densities vary with time and distance from the star. Choosing time $t$ as the variable, with increment $dt$, we have to numerically solve the following set of coupled difference equations in $r, V, T, n_{0}, n_{1}, n_{g}, \nu, a$, all functions of $t$:

$F=F(a)\,\,\, (eq. 11),\\$
$dV=\frac{GM_{*}(F-1)}{r^{2}(1-\frac{u_{H}^{2}}{V^{2}})}dt\,,\\$ 
$dr=Vdt\,,\\$
$T=T_{c}r_{c}/r\,,\\$
$u_{i}=(kT/m_{i})\,\, (i:H, monomer, grain)\,,\\$
$n{0}=n_{0c}\frac{V_{c}r_{c}^{2}}{V.r^{2}}\,,\\$
$dn_{1}=-c_{1}n_{1}n_{g}\nu^{2/3}dt\,,\\$
$n_{1}=n_{1c}\frac{V_{c}r_{c}^{2}}{V.r^{2}}+dn_{1}\,,\\$
$d\nu=(c_{1}n_{1}\nu^{2/3}+c_{1}n_{g}\nu^{7/6})dt\,,\\$
$n_{g}=(n_{0}-n_{1})/\nu\,,\\$
$a=10^{-8}\mu\nu^{1/3},$

where  $\sigma_{0}=9.1 \,10^{-16}$ cm$^{-2}$ or $6.1 \,10^{-16}$ cm$^{-2}$, the assumed cross-sections of SiC and carbonaceous monomers, respectively; $c_{1}=\sigma_{0}u_{1}$, and $u_{1}$ is the thermal velocity of a monomer. The cross-section of a grain is taken as $\sigma_{0}\nu^{2/3}$ for obvious reasons. As the mass of a grain is proportional to $\nu$, its thermal velocity is $u_{1}/\nu^{1/2}$. 

The initial conditions are defined by the time and radius of first condensation:

$t=0, r=r_{c}, V=u_{Hc}, T=T_{c},  n_{0}=n_{0c}, n_{1}=n_{1c}, n_{g}=n_{gc}$, and $\nu_{c}$ and $a_{c}$ determined by $F(a_{c})$=1.

Figure 4 illustrates this procedure in a typical case where the condensible material is carbon, with $R_{*}=2\,\,10^{13}$ cm, $L_{*}=2\,\,10^{4}\,\, L_{\odot}$, $\dot M=10^{-4}\,\,M_{\odot}$/y, $n_{Hc}=3.6\,\, 10^{11}$ cm$^{-3}$, $n_{gc}=10$\, cm$^{-3}$ at $r=r_{c}=4.2\,\,10^{13}$cm. We take the relative cosmic abundance of C to be $10^{-4}$, considering the large fraction captured into gaseous CO. Based on the colour temperatures of C-rich circumstellar shells (IRAS class 4n) displaying only weak or no 11.3-$\mu$m feature, we take the condensation temperature of carbonaceous grains to be 1200 K.

\begin{figure}
\resizebox{\hsize}{!}{\includegraphics{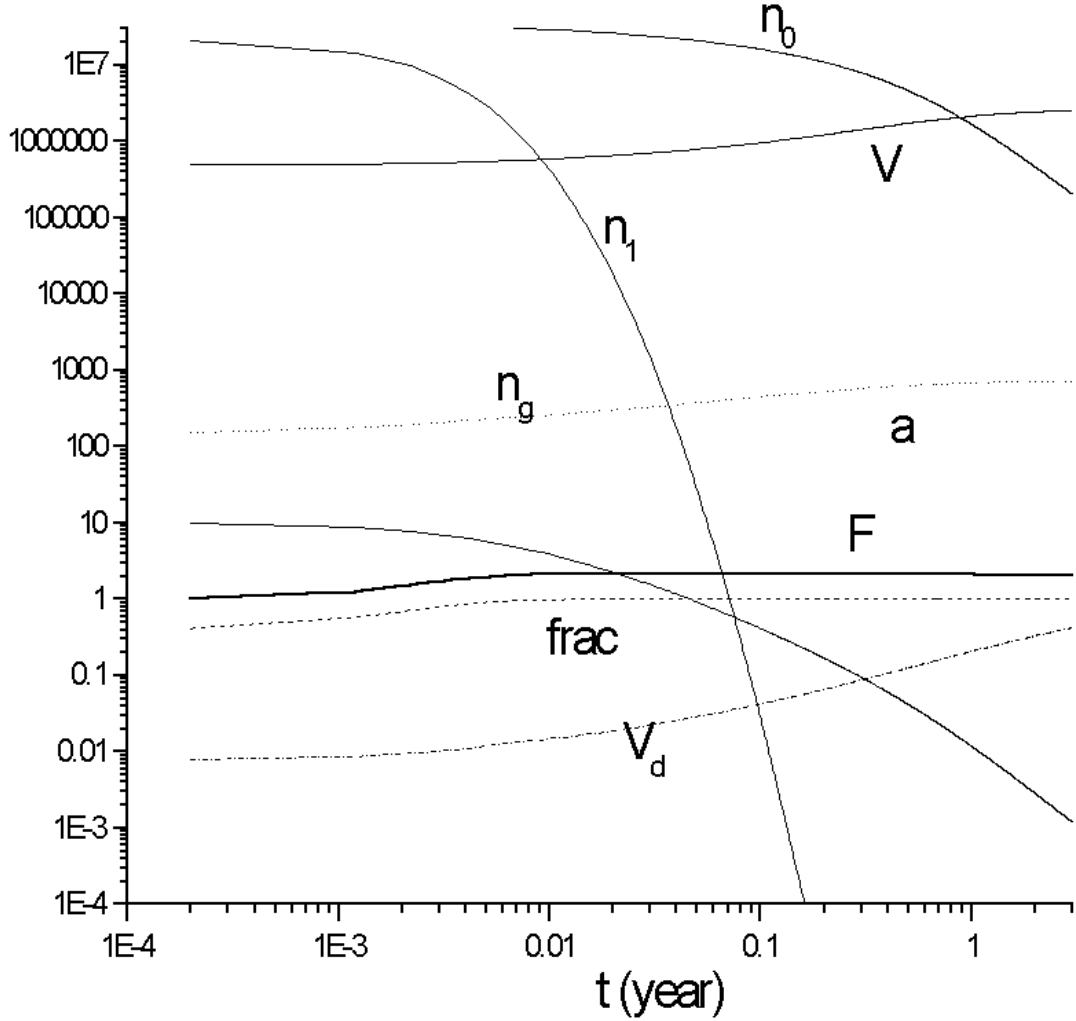}}
\caption[]{Expanding atmosphere; carbon is the sole condensible species; $L_{*}=2\,10^{4}\,L_{\odot}$ 
$\dot M=10^{-4}\,M_{\odot}/y$. Note the relative smallness of the drag velocity. Here the decrease of the grain number density is entirely due to expansion. $F_{max}=2.1$. Asymptotically, $V=25$ km.s$^{-1}$ and $a=0.0725 \,\mu$.} 
\end{figure}

Carbon grains are able to drive a stellar wind because they are quite opaque all over the spectrum, and because C is relatively abundant. Silicon is 10 times less abundant and relatively transparent at short wavelengths, so despite its early condensation, the maximum $F$ it can give rise to is generally smaller than 1, insufficient to lift the atmosphere.

\section{The expanding atmosphere; composite grains}

Even in cases where SiC by itself cannot drive a wind, the carbon condensing farther from the star may do so, as shown above, and drag SiC grains along. There is no particular difficulty in applying the dynamical equations of Sec. 3 to the case of such a composite atmosphere. There are now two sets of variables, denoted by 1 for SiC, and 2 for carbonaceous grains, governed by the same equations as above. We have to complement the previous system of equations with a description of the condensation of C atoms on themselves as well as on pre-existing SiC grains. Thus, the number of monomers carried by each type of grains now evolves according to

$d\nu_{1}=(c_{1}n_{1}\nu_{1}^{2/3}+c_{2}n_{2}\nu_{1}^{2/3}+c_{1}n_{g1}\nu_{1}^{7/6}+c_{1}n_{g2}\nu_{2}^{7/6})dt\,,\\$

$d\nu_{2}=(c_{2}n_{2}\nu_{2}^{2/3}+c_{2}n_{g2}\nu_{2}^{7/6})dt\,,\\$.

In order to avoid the case of a static atmosphere, we must assume, to begin with, that the conditions allow for a wind to be established, for instance due to stellar pulsations. This logically implies that the atmosphere begins to move up right from the star's photosphere. However, our simplified model cannot describe the dynamics below the sonic point for it assumes that the wind velocity there is null, or negligibly small (see Deguchi \cite {deg}). The process is therefore best followed in the time domain. In the case of Fig. 4, we saw that carbon condenses at an altitude of about $2\,\,10^{13}$ cm. In order to get a lower estimate of the time required for the gas from the photosphere to reach this altitude, assume its velocity to be uniformly sonic all along, say as high as $2\,\,10^{5}$ cm.s$^{-1}$. Even so, it would take 3 years for the gas to travel this vertical distance. We therefore consider that during, say, 10 years, SiC grains condense and grow in a nearly static atmosphere below the sonic point, as in Sec. 2. 

The calculations were carried out for a case where 
$R_{*}=2\,\,10^{13}$ cm, $L_{*}=10^{4}\,\, L_{\odot}$, $n_{Hc}=10^{11}$ cm$^{-3}$, $f_{at1}=10^{-5}$, $f_{at2}=2\,10^{-4}$, $n_{g1c}=10$ and $n_{g2c}=1$ cm$^{-3}$ at $r=r_{c}=4.2\,\,10^{13}$cm. The results are illustrated by Fig. 4.

SiC grains condense first (Fig. 5a). The Si reservoir has ample time to be depleted (frac(SiC)=1), and the grains to reach asymptotic size (0.0076 $\mu$m) before 10 y. Carbon remains gaseous except for the small fraction trapped in SiC. The force ratio $F$ (Fig. 5b) is distinctly smaller than 1, so the atmosphere cannot be lifted up at this stage.

The SiC grains cannot drive a wind because $F(a)$ remains low (cf. Sec. 2). A few years later, these grains have travelled farther from the star, and reached an altitude where the temperature is lower and carbon can condense, which it does both on existing SiC grains and on Si-free carbonaceous material. The SiC grains resume their growth in the form of a carbon coating, to reach a final asymptotic radius of 0.0154 $\mu$m. The parallel evolution of purely carbonaceous grains is seen to end up in grains of 0.0308 $\mu$m, including most of the initially available gaseous carbon. 

As the total dragging force is now the sum of the forces on both types of grains, as defined by eq. 11, the ratio of radiative to gravity forces quickly increases as soon as carbon begins to condense, say after 10 y,  and grows beyond 1 (fig. 5b) so the atmosphere is indeed accelerated upwards to reach an asymptotic velocity of 13 km.s$^{-1}$.

The SiC grains are seen, in Fig. 5a, to reach an asymptotic size of 0.0154 $\mu$ within 1 y or so, due to the formation of a carbon coating. Here, the volume ratio of coating to core is about 7.3. According to the calculations of Kozasa et al. \cite{koz96}, this is large enough to completely screen the SiC 11.3-$\mu$ feature, and explain its disappearance from the stellar spectrum. The latter then reduces to the cold continuum of carbonaceous grains, as well as from the interstellar medium, even though it certainly harbours the expelled SiC grains. 

Figures 5a also shows the formation of carbon grains 0.0308 $\mu$m in radius carrying 63$\%$ of the initially available carbon. The warm emission continuum of this circumstellar material is therefore expected to override the hotter photospheric spectrum. Note that the final dust-to-gas mass ratio reached an asymptotic value of $3.8\,\,10^{-3}$.

\begin{figure}
\resizebox{15 cm}{18 cm}{\includegraphics{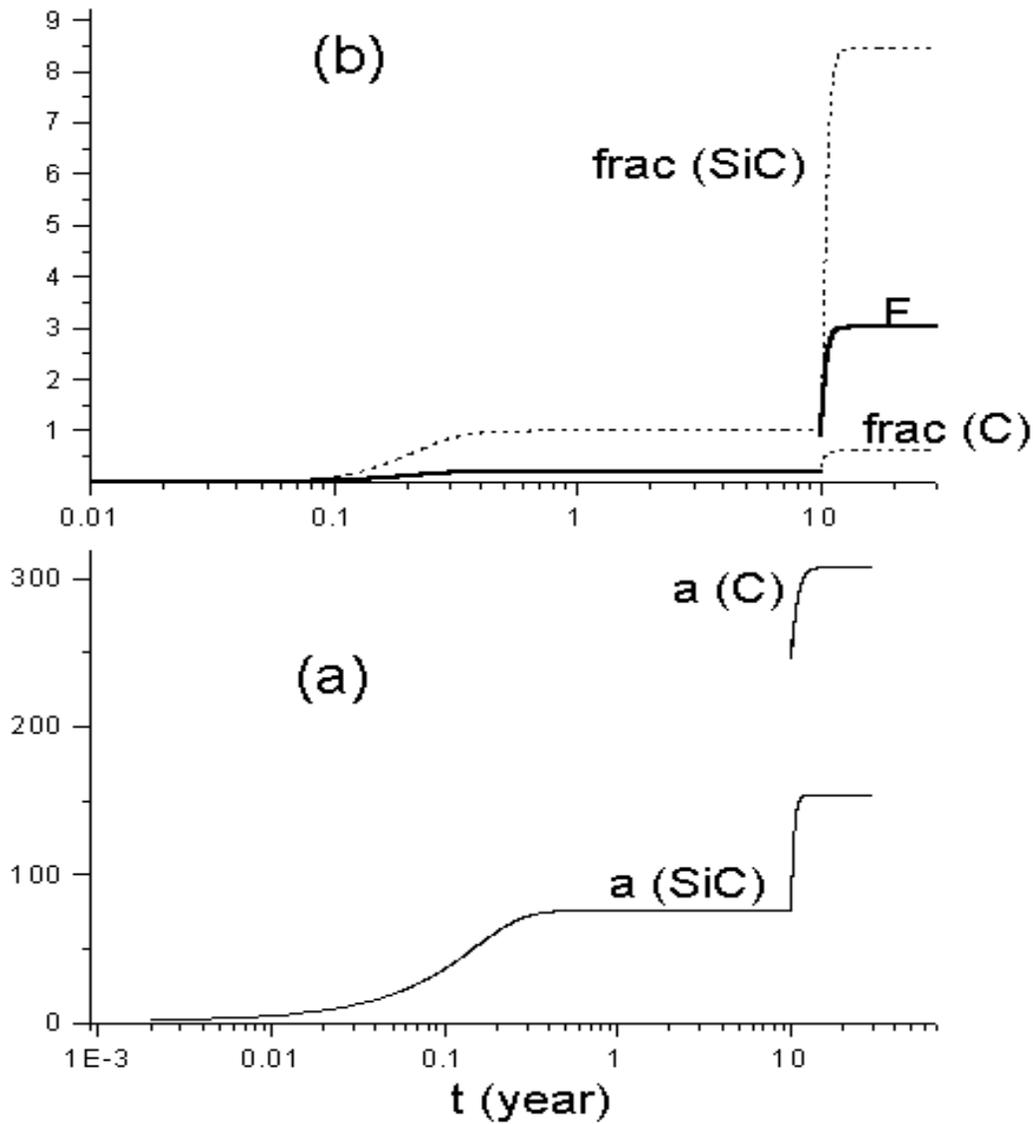}}
\caption[]{Expanding atmosphere; a) grain radii a(SiC) and a(C), in Angstroms; b) the total force ratio $F$, and the fractions of material condensed in SiC grains (frac(SiC), including the carbon mantle) and in carbonaceous grains (frac(C)); see comments in the text.}
\end{figure}

 These results are sensitive to the parameters, essentially cosmic abundances, stellar parameters and  initial number densities of grains, so a large gamut of mantle-to-core volume ratios is to be expected, with a corresponding range of the screening effect on the SiC feature.
Thus, increasing the number density of carbonaceous seeds from 1 to 10 cm$^{-3}$ decreases the mantle-to-core volume ratio from 7.3 to 3.8, while the final fraction of carbon trapped in carbonaceous grains increases from 0.63 to 0.81.

\section{Conclusion}
The argument developed above provides an explanation of two extreme situations regarding the 11.3-$\mu$m SiC feature: a) the case of the static atmosphere of a C-rich star, where condensed SiC gives rise to a strong emission feature protruding from the hot photospheric continuum; b) the case of an expanding atmosphere, where enough carbon can condense upon pure SiC grains to form a thick mantle which completely shields the feature behind a  warm continuum, so that it is not even observed in the interstellar medium. 

The great variety of intermediate stellar spectra as documented in Baron et al. \cite{bar}, where the contrast between feature and continuum roughly spans the range 0.1 to 3 can be interpreted in terms of both static and expanding atmospheres.

Indeed, it must be remembered that some phenomenon (stellar pulsations ?) is required to lift the SiC grains up to altitudes where the temperature is low enough, but the density still large enough, for substantial carbon to condense, otherwise the atmosphere remains static as in Sec. 2. Hence, intermediate cases may be contemplated, where the required process is not (yet) complete or steady, so carbon condensation remains only partial and distributed over various altitudes, with different densities and temperatures. Even if the wind is steady, a continuous distribution of mantle-to-core volume ratios may result from different initial number densities of seeds, a highly erratic parameter. This may explain the regular trend observed by Baron et al. \cite {bar} over the 538 IRAS 4n spectra: as the 11.3-$\mu$m SiC feature is progressively buried into the carbonaceous continuum, the slope of the latter indicates a decreasing temperature, characteristic of carbonaceous grains in thicker and upper layers of the circumstellar shell.

\end{document}